\documentclass[aps,amsmath,twocolumn,superscriptaddress,prl]{revtex4}
\usepackage{amsmath,amssymb,bm}
\usepackage[dvips]{graphicx}
\usepackage{braket}
\usepackage{ulem}
\usepackage{setspace}[10]
\usepackage{comment}
\usepackage[dvips]{color}
\usepackage[dvipsnames]{xcolor}
\usepackage{comment}
\usepackage{mathrsfs}

\begin{document}
\title{Magnon Pairs and Spin-Nematic Correlation in the Spin-Seebeck Effect}

\author{Daichi Hirobe}
\email{daichihirobe@ims.ac.jp}
%\thanks{These two authors contributed equally.}
\affiliation{Institute for Materials Research, Tohoku University, Sendai 980-8577, Japan}

\author{Masahiro Sato}
\email{masahiro.sato.phys@vc.ibaraki.ac.jp}
%\thanks{These two authors contributed equally.}
\affiliation{Spin Quantum Rectification Project, ERATO, Japan Science and Technology Agency, Sendai 980-8577, Japan}
\affiliation{Department of Physics, Ibaraki University, Mito, Ibaraki 310-8512, Japan}

\author{Masato Hagihala}
%\email{}
%\thanks{These two authors contributed equally.}
\affiliation{Institute of Solid State Physics, The University of Tokyo, Kashiwa, Chiba 277-8581 Japan}
\affiliation{Institute of Materials Structure Science, High Energy Accelerator Research Organization (KEK), Tokai, Ibaraki 319-1106, Japan}

\author{Yuki Shiomi}
\affiliation{Department of Basic Science, University of Tokyo, Meguro, Tokyo 153-8902, Japan}

\author{Takatsugu Masuda}
%\email{masuda@issp.u-tokyo.ac.jp}
%\thanks{These two authors contributed equally.}
\affiliation{Institute of Solid State Physics, The University of Tokyo, Kashiwa, Chiba 277-8581 Japan}

\author{Eiji Saitoh}
%\email{eizi@imr.tohoku.ac.jp}
\affiliation{Institute for Materials Research, Tohoku University, Sendai 980-8577, Japan}
\affiliation{Spin Quantum Rectification Project, ERATO, Japan Science and Technology Agency, Sendai 980-8577, Japan}
\affiliation{The Advanced Science Research Center, Japan Atomic Energy Agency, Tokai 319-1195, Japan}
\affiliation{WPI Advanced Institute for Materials Research, Tohoku University, Sendai 980-8577, Japan}

\date{\today}

%%%%%%%%%%%%%%%%%%%%%%%%%%%%%%%%%%%%%%%%%%%
%%%%%%%%%%%%%%%%%%%%%%%%%%%%%%%%%%%%%%%%%%%
%%%%%%%%%%%%%%%%%%%%%%%%%%%%%%%%%%%%%%%%%%%

\begin{abstract}

	\textcolor{black}{Investigating exotic magnetic materials with spintronic techniques 
	is effective at advancing magnetism as well as spintronics.
	In this work,} we report unusual field-induced suppression of the spin-Seebeck effect (SSE) in 
	a quasi one-dimensional frustrated spin-$\frac{1}{2}$ magnet LiCuVO$_4$, 
	known to exhibit spin-nematic correlation in a wide range of external magnetic field $B$. 
	The suppression takes place above $|B| \agt 2$ T 
	in spite of the $B$-linear isothermal magnetization curves in the same $B$ range.
	\textcolor{black}{The} result \textcolor{black}{can be} attributed to the growth 
	of the spin-nematic correlation while increasing $B$.
	%Instead, our result is attributed to the growth of the spin-nematic correlation with an increase of $B$.
	The correlation stabilizes magnon pairs carrying spin-2, thereby suppressing
	the interfacial spin injection of SSE by preventing the spin-1 exchange between single magnons 
	and conduction electrons at the interface. 
	This interpretation is supported by integrating thermodynamic measurements and 
	theoretical analysis \textcolor{black}{on} the SSE.
\end{abstract}

%%%%%%%%%%%%%%%%%%%%%%%%%%%%%%%%%%%%%%%%%%%
%%%%%%%%%%%%%%%%%%%%%%%%%%%%%%%%%%%%%%%%%%%
%%%%%%%%%%%%%%%%%%%%%%%%%%%%%%%%%%%%%%%%%%%
\maketitle

%%%%%%%%%%%%%%%%%%%%%%%%%%%%%%%%%%%%%%%%%%%
%%%%%%%%%%%%%%%%%%%%%%%%%%%%%%%%%%%%%%%%%%%
%%%%%%%%%%%%%%%%%%%%%%%%%%%%%%%%%%%%%%%%%%%
%\noindent
\textit{Introduction}$-$
	\textcolor{black}{Spin-Seebeck effects} 
	(SSE)~\cite{Uchida2010, Xiao, Adachi, Zhang, Hoffman, Rezende, Wu2015, Kikkawa2015, Jin, Kehlberger, 
	Aqeel, Seki, Wu2016, Geprags, Li, Hirobe2017, Hirobe2018, Liu, Qiu2018} 
	refer to the generation of a spin current owing to a temperature gradient 
	in a magnetic material. It takes place in a magnetic insulator with a metallic contact.
	When nonequilibrium magnons are accumulated \textcolor{black}{at} the interface due to
	a temperature gradient,
	the annihilation of such a single magnon is followed by the flip of a conduction-electron spin 
	via the interfacial exchange interaction. 
	As a result, the exchange of spin-1 takes place dominantly, enabling conversion from a magnon spin current 
	into a conduction-electron one~\cite{Adachi}. The latter spin current can be detected as a transverse electric field 
	via the inverse spin-Hall effect~\cite{Azevedo, Saitoh, Valenzuela, Sinova} in the metallic contact.
	SSEs have been found to take place even \textcolor{black}{in} paramagnetlike insulators 
	with spin correlations~\cite{Hirobe2017, Hirobe2018, Liu}. 
	These findings point to the use of SSE as a probe for spin correlations without the magnetic orders,
	\textcolor{black}{for example, in quantum spin systems~\cite{XGWen, Sachdev, Balents, Giamarchi}}.
	
	%Collective spin excitations different from magnons can emerge, especially in quantum spin systems. 
	%These spin systems possess frustration or quantum fluctuation enhanced by the low-dimensionality 
	%and the smallness of the localized spins, often spin-$\frac{1}{2}$~\cite{XGWen, Sachdev, Balents}.
	%As a result, conventional magnetic correlations, that is, magnetic dipolar correlations, and 
	%their corresponding magnons are \textcolor{black}{well} suppressed.
	%Instead, higher-rank magnetic multipolar correlations and corresponding exotic quasi-particles 
	%can emerge {\textcolor{black}{in} quantum spin systems. 
	
	The magnetic quadpolar correlation, also known as the spin-nematic correlation~\cite{Penc,Shannon,Chubukov},
	is the simplest example of magnetic multipolar correlations. 
	It represents the correlation between magnon pairs, rather than single magnons. 
	To stress this point, the spin-nematic correlation will be called the magnon-pair correlation 
	hereafter.
	%The simplest example is a magnetic quadpolar correlation, 
	%also known as a spin-nematic correlation. The correlation represents the one between 
	%magnon pairs, rather than single magnons. 
	\textcolor{black}{A typical} magnon-pair correlation appears in a one-dimensional (1D) frustrated 
	spin-$\frac{1}{2}$ chain with the ferromagnetic nearest neighboring exchange interaction $J_1 < 0$ and
	the antiferromagnetic next nearest neighboring one $J_2 > 0$. 
	The Hamiltonian of this 1D $J_1$-$J_2$ model reads
	\begin{eqnarray}
	{\cal H}=\sum_{j}\left(J_1{\bm S}_j\cdot{\bm S}_{j+1}+J_2 {\bm S}_j\cdot{\bm S}_{j+2}-g\mu_B BS_j^z\right).
	\label{eq:J1-J2}
	\end{eqnarray}
	Here ${\bm S}_j$ is the spin-$\frac{1}{2}$ operator on the $j$th site, and the site number $j$ increases 
	along the spin-chain direction. The last term represents the Zeeman interaction with 
	the external magnetic field $B$ \textcolor{black}{along the $z$-axis} with $g$ and $\mu_B$ being respectively the g-factor and the Bohr magneton. 
	The low-energy physical properties of Eq. (\ref{eq:J1-J2}) and its variants 
    have been elucidated~\cite{Vekua,Hikihara,Sudan,Sato09,Sato11,Sato13} 
	using powerful theoretical techniques in the last decade.
	The ground-state diagram with $|J_1/J_2| = {\cal O}(1)$~\cite{Hikihara,Sudan} is schematically shown in Fig. 1(a) 
	as a function of $B$.
	In the lower $B$ range, a Tomonaga-Luttinger liquid (TLL)~\cite{Giamarchi} 
    with a vector spin chirality~\cite{Nersesyan,Hikihara01,Furukawa10,Furukawa12,Sato-Sup} appears.
	As $B$ is increased, the magnon-pair correlation grows to give rise to a spin-nematic 
	TLL~\cite{Vekua,Hikihara,Sudan,Sato09,Sato11,Sato13} in a wide $B$ range. 
	In this state, single magnons acquire an energy gap equivalent to the binding energy 
	of magnon pairs while magnon pairs are gapless. Accordingly, a change in spin angular momentum is 
	quantized in units of $2\hbar$, not $\hbar$, in low energy.
	
	In this \textcolor{black}{study}, we have investigated the SSE \textcolor{black}{in an} 
	insulating quantum magnet LiCuVO$_4$~\cite{Enderle, Naito, Buttgen2012, Svistov}.
	LiCuVO$_4$ is an established model material \textcolor{black}{for} a strong magnon-pair correlation, 
	representing a family of quasi-1D $J_1$-$J_2$ 
    magnets~\cite{Hase,Goto,Wolter,Wolter2,Nawa2014,Nawa2017,Dutton2012,Grafe2017,Arcon}. 
	\textcolor{black}{
	Since the spin quantum number carried by quasiparticles is increased effectively by  magnon-pair formation, 
	the SSE seems to be enhanced while increasing $B$.  
	Contrary to this na\"ive expectation,} 
	the SSE \textcolor{black}{in} LiCuVO$_4$ has been observed to exhibit a strong $B$-induced suppression 
	\textcolor{black}{alongside} the $B$-linear magnetization curves above the magnetic ordering temperatures. 
	Such a $B$ response of the SSE is different from those of 
	magnetically ordered states~\cite{Uchida2010,Seki,Wu2016,Geprags} and a 1D quantum spin liquid~\cite{Hirobe2017}.
	We interpret the result as the evidence for $B$-induced crossover from the single-magnon 
	correlation to the magnon-pair one, and its resulting prevention of the interfacial exchange 
	of spin-1 in the SSE. 
	Observing the magnon-pair correlation is generally difficult. 
	Our study shows that SSE serves as a powerful probe for dynamical and transport natures of 
    such spin-nematic states in quantum magnets.

%%%%%%%%%%%%%%%%%%%%%%%%%%%%%%%%%%%%%%%%%%%
%%%%%%%%%%%%%%%%%%%%%%%%%%%%%%%%%%%%%%%%%%%
%%%%%%%%%%%%%%%%%%%%%%%%%%%%%%%%%%%%%%%%%%%
%\noindent
\textit{Spin-nematic nature of LiCuVO$_4$}$-$
	LiCuVO$_4$ is a typical Mott insulator for which experimental evidences for 
	the magnon-pair correlation have been established. 
	A spin chain embedded in LiCuVO$_4$ is shown in Fig. 1(b). 
	Each Cu$^{2+}$ ion carries spin-$\frac{1}{2}$ and they form a 1D chain 
	along the $b$-axis by sharing O$^{2-}$ ions. 
	If the weak inter-chain interaction $J'$ is \textcolor{black}{ignored}, 
	LiCuVO$_4$ can be \textcolor{black}{well} described by Eq. (\ref{eq:J1-J2}).
	The magnitudes of $J_1$ and $J_2$ were estimated experimentally, for example, 
	from neutron scattering spectra~\cite{Enderle,Svistov,Enderle2,Drechsler2011}: 
	$J_2 = 40\sim70$ K and $|J_1 / J_2| = {\cal O}(1)$. 
	Because of the weak $J'$ in LiCuVO$_4$, magnetically ordered phases appear at low temperatures; 
	however, each phase nicely reflects the phase diagram of the purely 1D model 
	[see also Fig. 1(a)]. 
	$J'$ was estimated experimentally to be a few Kelvin~\cite{Enderle,Svistov}, 
	consistent with the magnetic ordering temperatures ($T_c$) of about 3 K~\cite{Enderle,Svistov,Naito,Buttgen2012}. 
	In a low-$B$ range below $T_c$, a spin spiral order appears~\cite{Naito,Buttgen2012,Furukawa10}, 
	reflecting the TLL with a vector spin chirality~\cite{Furukawa10,Furukawa12,Sato-Sup}. 
	As $B$ is increased to about 7 T, a spin-density-wave (SDW) order appears as shown in Fig. 1(c), 
	and it continues up to about 50 T ($\sim J_{2}$). 
	Immediately below the saturation magnetization, a three-dimensional (3D)
	spin-nematic order may occur~\cite{Orlova,Sato13,Zhitomirsky,Ueda}, whose possible existence attracts attention. 
	Importantly, the magnon-pair (spin-nematic) correlation evidently persists above $T_c$,
	and exhibits a quasi long-range order over a wide $B$ range 
	together with the SDW correlation.
	In Refs.~\onlinecite{Sato09} and \onlinecite{Sato11}, 
	a theoretical proposal was made for detecting signs of the spin-nematic TLL by 
	nuclear magnetic resonance (NMR) and neutron scattering techniques, 
	followed by experimental observations of those signs~\cite{Nawa13, Nawa14, Masuda, Mourigal}.
	%\tr{The ways of detecting signatures of the spin-nematic TLL has been proposed 
	%in Refs.~\onlinecite{Sato09,Sato11}, and actually, those have been observed experimentally 
	%by NMR~\cite{Nawa13, Nawa14} and neutron scattering~\cite{Masuda, Mourigal} studies.}
%neutron scattering~\cite{Masuda, Mourigal} studies}	
%This defining characteristic has been observed 
%experimentally by NMR~\cite{Nawa13, Nawa14} and 
%neutron scattering~\cite{Masuda, Mourigal} studies, and can be described 
%in the form of the spin-nematic Tomonaga-Luttinger liquid.

	%\twocolumngrid
	%\onecolumngrid
	\begin{figure}[th]
	\begin{center}
	\includegraphics{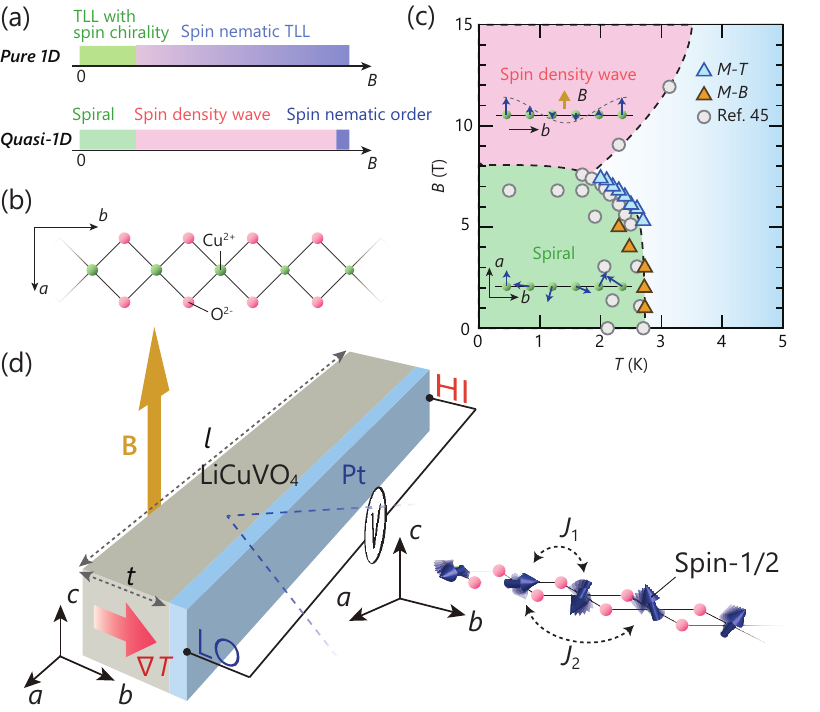}
	\end{center}
	\caption{
	(a) Theoretical ground-state phase diagrams of a purely 1D frustrated $J_1$-$J_2$ 
    spin-$\frac{1}{2}$ chain~\cite{Hikihara,Sudan} (top) and a quasi-1D one 
    with an interchain exchange interaction~\cite{Sato13} (bottom). 
	$B$ denotes external magnetic field. %TLL stands for Tomonaga-Luttinger liquid.
	(b) Spin chain in LiCuVO$_4$ composed of Cu$^{2+}$ and O$^{2-}$ ions.
	(c) Magnetic field ($B$) $-$ temperature ($T$) phase diagram of LiCuVO$_4$,
	obtained while applying $B$ in the $c$-axis.
	Triangular data points were taken in this study: the sky-blue ones from 
	the $T$ dependence of the magnetization $M$; the orange ones from the $B$
	dependence of $M$. The circular data points were adapted from Ref.~\onlinecite{Buttgen2012}.
	(d) Experimental set-up for detecting the spin-Seebeck effect in a LiCuVO$_4$/Pt system.
	$J_1$ and $J_2$ respectively denote the nearest and next nearest neighboring exchange interactions
	in the spin chain of LiCuVO$_4$; $\nabla T$ a temperature gradient along the spin chain;
	$t$ and $l$ respectively the thickness and the length of the LiCuVO$_4$.
	%\textcolor{red}{\textit{reduced by 26} The direction of $B$, that is, the $c$-axis corresponds to the $z$-axis 
	%in Eq. (\ref{eq:J1-J2}) while the $a$- and $b$-axes to the $x$- and $y$-axes, respectively.}
	%(e), (f) $T$ dependence of $M$ at $B=1$ T (e) and ${\rm d}(\chi T)/{\rm d}T$ 
	%with $\chi = M/B$ at several $B$ (f). The inset to (e) is a magnified view below 30 K.
	%(g), (h) $B$ dependence of $M$ (g) and ${\rm d}M/{\rm d}B$ (h) at several $T$. 
	}
	\end{figure}
	%\twocolumngrid

%%%%%%%%%%%%%%%%%%%%%%%%%%%%%%%%%%%%%%%%%%%%
%%%%%%%%%%%%%%%%%%%%%%%%%%%%%%%%%%%%%%%%%%%%
%%%%%%%%%%%%%%%%%%%%%%%%%%%%%%%%%%%%%%%%%%%%
%\noindent
\textit{Experimental details}$-$
   Single crystals of LiCuVO$_4$ were grown by a travelling-solvent floating-zone method,
   which was exactly the same as reported one of the present authors~\cite{Masuda}.
   The grown single crystals were cut into cuboids that were typically 5 mm along the $a$-axis 
   and 1 mm along the $b,c$-axes for SSE measurements.  
   %\textcolor{red}{
   %   \textit{reduced by 40 words}
   %   The magnetization vs temperature of the single crystals was measured with the Magnetic Property
   %   Measurement System (Quantum Design, Inc.) while the isothermal magnetization curves 
   %   were taken with a vibrating sample magnetometer for the Physical Property Measurement System 
   %   (Quantum Design, Inc.).}
   Temperature ($T$) and magnetic field ($B$) dependences of the magnetization were 
   found to be consistent with a $B$$-$$T$ phase diagram reported elsewhere~\cite{Buttgen2012}, 
   as shown in Fig. 1(c). 
   \textcolor{black}{
   	%\textit{increased by 4 words}
   The experimental details and the magnetic properties are described in Supplemental Material~\cite{SuppMater}.
   }

   We used a LiCuVO$_4$/Pt junction system as shown in Fig. 1(d) 
   to investigate the SSE. 
   %\textcolor{red}{
   %   \textit{reduced by 50 words}
   %   The surface of LiCuVO$_4$
   %   perpendicular to the spin-chain direction (the $b$ axis) was polished mechanically, and 
   %   the 7-nm-thick Pt film was sputtered on the polished surface in an Ar atmosphere. 
   %   The sample was sandwiched between sapphire plates, and the rear of the LiCuVO$_4$ was 
   %   thermally anchored on a heat sink.}
   A temperature gradient $\nabla T$ was applied along the spin chains \textcolor{black}{with a heater}.
   \textcolor{black}{We created} the temperature difference $\Delta T$ 
   between the top of the Pt film and the rear of the LiCuVO$_4$. 
   %\textcolor{red}{
   %   \textit{reduced by 14}
   %   using a chip resistance heater on the sapphire plate attached to the Pt film.}
   Au wires were attached to the ends of the Pt film to obtain the DC voltage $V$, 
   \textcolor{black}{for which we excluded a background voltage signal taken with the heater off}. 
   The magnetic field $B$ was applied \textcolor{black}{along the $c$-axis, being} 
   perpendicular both to $\nabla T$ and the direction 
   across the electrodes\textcolor{black}{; thus, the $c$-axis corresponds to the $z$-axis
   in Eq. (\ref{eq:J1-J2}) while the $a$- and $b$-axes to the $x$- and $y$-axes, respectively.} 
   To quantitatively compare the voltage signals, we show 
   the transverse thermopower $S=j_e/|\nabla T|\approx(V/\Delta T \rho)(t/l)$. 
   Here $j_e$ is the current density in the Pt film due to thermoelectric effects, 
   $\rho$ is the electrical resistivity of the Pt film, and 
   $t$ and $l$ are respectively the thickness and the length of the LiCuVO$_4$. 
   Additionally, we defined the average temperature $T_\mathrm{ave}$ as 
   $T_\mathrm{ave}=(T_\mathrm{H} + T_\mathrm{L})/2$
   in which $T_\mathrm{H}=T+\Delta T$ and $T_\mathrm{L}=T$ are respectively the temperatures of
   the top  of the Pt film and the rear of the LiCuVO$_4$.
   %\textcolor{red}{\textit{reduced by 14}
   %All voltage signals were taken in the Physical Property Measurement System (Quantum Design, Inc.).}
   \textcolor{black}{%\textit{increased by 11}
   The experimental details of SSE measurements are described in Supplemental Material~\cite{SuppMater}.}
   
%%%%%%%%%%%%%%%%%%%%%%%%%%%%%%%%%%%%%%%%%%%%
%%%%%%%%%%%%%%%%%%%%%%%%%%%%%%%%%%%%%%%%%%%%
%%%%%%%%%%%%%%%%%%%%%%%%%%%%%%%%%%%%%%%%%%%%
%\noindent
\textit{Experimental results for SSE}$-$
	\begin{figure}[t]
	%\begin{center}
	\includegraphics{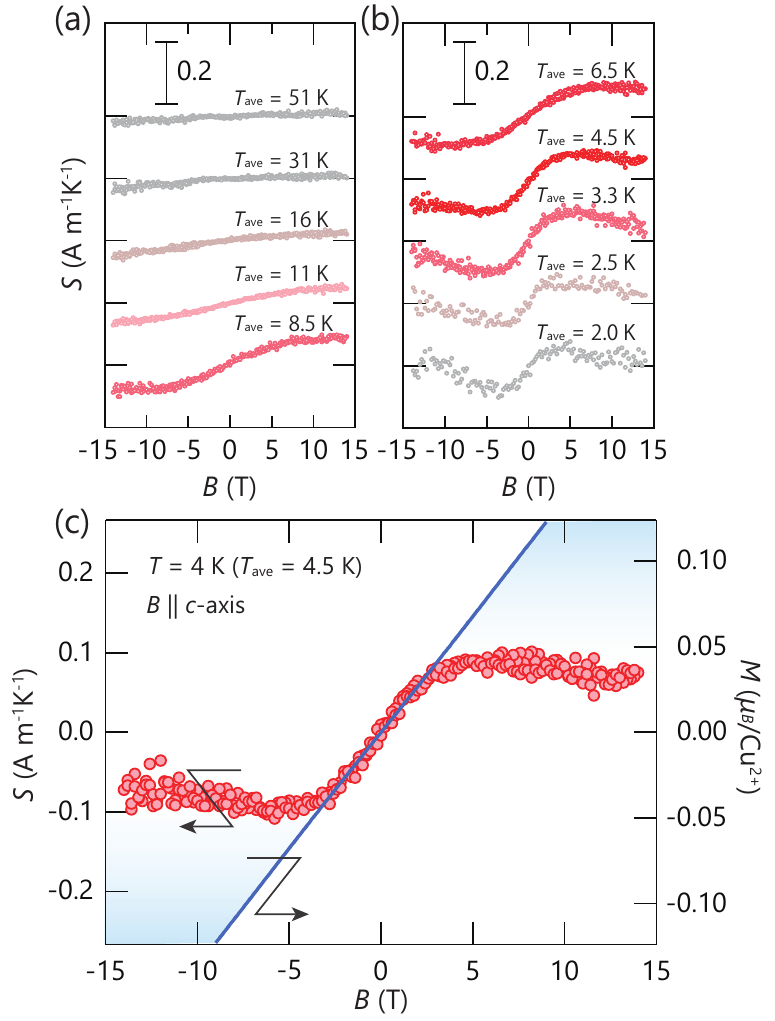}
	%\end{center}
	\caption{
	(a), (b) $B$ dependence of the transverse thermopower $S$ at several $T$.
	(c) Comparison between $B$ dependences of $S$ and $M$ at $T=4$ K. 
	}
	\end{figure}	
	In Figs. 2(a) and (b), we show the $B$ dependence of 
	the transverse thermopower $S$ at several $T_\mathrm{ave}$.
	A small $S$ was detected at 51 K, and found to be $B$-linear. 
	This \textcolor{black}{can} be explained by	
	the normal Nernst effect of Pt~\cite{Wu2015, Hirobe2017}. 
	However, as $T_\mathrm{ave}$ is decreased down to 11 K, a clear signal appears. 
	Its sign reverses when the magnetization is reversed, 
	which is a typical feature of SSE. Interestingly, $S$ starts 
	deviating from a $B$-linear line, and decreases while increasing $B$.
	As shown in Fig. 2(b), the deviation enhances with a further decrease of 
	$T_\mathrm{ave}$ down to 2 K, the lowest temperature in this study.
	
	To look into this $B$-dependence of $S$ in more detail, 
	we compare the $B$ dependences of 
	$S$ and the magnetization $M$ at $T=4$ K in Fig. 2(c). 
	Remarkably, in spite of the $B$-linear \textcolor{black}{change in} $M$, 
	$S$ gets suppressed strongly 
	\textcolor{black}{while increasing} $B$, and even exhibits a negative slope at $|B|\agt 5$ T. 
	%The suppression enhances with a further decrease of 
	%$T$ down to 2 K, the lowest temperature in this study.
	We stress that the suppression of $S$ cannot be attributed 
	to magnetic phase transitions since it takes place even above 
	$T_\mathrm{c}$ [see also Fig. 1(c)]. 
	Additionally, the Zeeman energy gap in spin excitations is unlikely to explain 
	the $B$-induced suppression of $S$ although seemingly similar results 
	were reported for ferrimagnets and paramagnets~\cite{Kikkawa2015, Jin, Liu}. 
	\textcolor{black}{%\textit{increased by 36}
	Generally the Zeeman energy gap starts suppressing thermal magnetic excitations 
	as the magnetization approaches saturation at low temperatures.
	Since $M$ of LiCuVO$_4$ is $B$-linear alongside $\sim 0.1$ $\mu_\mathrm{B}$/Cu$^{2+}$ even at $B=9$ T, 
	the smooth $M$$-$$B$ curve indicates the existence of a gapless 
	magnetic excitation~\cite{OYA1997, Oshikawa2000, Hastings2004}.
	%Therefore, thermal spin excitations persist within the present $B$ range.
	}
	%\textcolor{red}{\textit{reduced by 80}
	%In these previous cases, 
	%the $B$-induced suppression of SSEs accompanies the almost saturated magnetization.
	%This means that the Zeeman energy gap is wide enough to suppress thermal spin excitations, 
	%thereby suppressing SSEs. In the present case, however, $M$ is $B$-linear along with 
	%$\sim 0.1$ $\mu_\mathrm{B}$/Cu$^{2+}$ even at $B=9$ T, well below the saturation magnetization. 
	%Therefore, thermal spin excitations should still survive within the present $B$ range. 
	%This can rule out the Zeeman energy gap as a cause of the suppression of $S$.}
	
	The unusual suppression of $S$ invokes the magnon-pair correlation,
	which yields magnon pairs with a binding energy $E_\text{bind}$. 
	$E_\text{bind}$ has been predicted to already exist near zero magnetic field~\cite{Sato13}.
	Figure 3(a) shows the calculated $B$ dependence of $E_\text{bind}$
	for a purely 1D case with $|J_1/J_2| = 1$~\cite{Sato13}. 
	$E_\text{bind}$ increases linearly with $B$ alongside the $B$-linear 
	magnetization when $B$ is much lower than the saturation field [see also the inset to Fig. 3(a)].
	Within this framework, the $B$-induced $E_\text{bind}$ stabilizes magnon pairs 
	while inhibiting thermal excitation of single magnons.
	Because spin injection of SSE at the interface stems mainly from the exchange of spin-1, 
	spin-2 magnon pairs cannot contribute to such spin injection,
	thereby decreasing SSE signals.
	\textcolor{black}{
	The ability to selectively probe spin-1 magnetic excitations 
	should differentiate SSE measurements from thermal conductivity measurements.
	This is because the latter measurements simultaneously probe phonons as well as 
	multiple magnetic excitations carrying spin-1 and spin-2~\cite{SuppMater}.
	%In general selectively detecting spin-1 magnetic excitations is 
	%difficult in other transport measurements. 
	%With thermal conductivity measurements, 
	%We tried to find a possible relation between the magnon thermal conductivity and the energy gap. 
	%However, we found it difficult to deconvolute components 
	%of the measured thermal conductivity, because phonons and 
	%various magnetic excitations can contribute to heat conduction in LiCuVO$_4$~\cite{SuppMater}.
	}

%%%%%%%%%%%%%%%%%%%%%%%%%%%%%%%%%%%%%%%%%%%%
%%%%%%%%%%%%%%%%%%%%%%%%%%%%%%%%%%%%%%%%%%%%
%%%%%%%%%%%%%%%%%%%%%%%%%%%%%%%%%%%%%%%%%%%%
%\noindent
\textit{Comparison between experimental and theoretical results}$-$
%	We are in a position to discuss the $B$-induced crossover
%	from the single-magnon correlation to the magnon-pair one. 
%	This is done by calculating interfacial spin currents, $\tilde{J}_\text{s}$, 
%	injected from a magnet into a metal on the basis of a microscopic SSE formalism 
%	(see Supplemental Material). 
	We theoretically calculate spin currents injected from a magnet (LiCuVO$_4$) to a metal (Pt)
	and compare them with $S$, because inverse spin-Hall voltages are proportional to injected spin currents.
	For simplicity, we assume that the spin dynamics of LiCuVO$_4$ is described by a spin-nematic TLL, 
	ignoring the weak inter-chain interactions.
	We also make the conventional assumption that a weak exchange interaction $J_{\text{sd}}$ exists at the interface 
	between the magnet and the metal. 
	The normalized spin current $\tilde{J}_\text{s}$~\cite{Jauho,Adachi,Hirobe2017} is then given by 
	(see Supplemental Material)
    %To support the above argument about the suppression in the SSE, we will compare the experiments 
    %with the theoretical result of the tunnel spin current between the magnet (LiCuVO$_4$) and the metal (Pt), 
    %which would be proportional to $S$.
    %In the computation, for the simplicity, we assume that the spin dynamics of LiCuVO$_4$ is described 
    %by a spin-nematic TLL and the weak interaction coupling is negligible. 
    %We also take the reasonable assumption that a weak exchange interaction $J_{\text{sd}}$ exists 
    %at the interface between LiCuVO$_4$ and Pt. The renormalized tunnel spin current 
    %$\tilde{J}_\text{s}$~\cite{Jauho,Adachi,Hirobe2017} is given by (see Supplemental Material)
	\begin{eqnarray}
	\tilde J_s = \frac{1}{T^{2}} \int d\omega\,\, 
	{\rm Im}\chi_{\rm mag}^{-+}(\omega, T)\,\,
	\frac{\omega^2}{1+\tau_s^2\omega^2}\,\,
	\frac{1}{\sinh^2(\omega/(2T))},
	\label{eq_SpinCurrent2}
	\end{eqnarray}
    up to the leading order of $J_{\text{sd}}$. 
    Here, $\omega$ is the angular frequency, $T$ is the mean value of the two temperatures of the magnet and the metal,
    and $\tau_s$ is the spin relaxation time for the metal. The integral range is $(-\infty, \infty)$.
	%Here, $\omega$ is the angular frequency, 
	%the integral range is $(-\infty, \infty)$, 
    %$T$ ($k_B=1$) is the averaged temperature of the magnet and the metal, 
    %and $\tau_s$ the spin relaxation time for the metal.
	$\chi_{\rm mag}^{-+}$ denotes the dynamical spin susceptibility of the magnet, 
	and describes \textcolor{black}{%\textit{increased by 11} 
	the dynamics of a single magnon 
	(strictly speaking, a paramagnon in a spin-nematic TLL).} 
    In formula~(\ref{eq_SpinCurrent2}), %$\tilde{J}_\text{s}$ 
    the spin current is injected by single magnons 
    which have an energy gap due to magnon-pair formation. 
    %and we ignore the magnon-pair driven spin current considering its small contribution~\cite{SuppMater}. 
	We have ignored the magnon-pair-driven spin current considering its small magnitude~\cite{SuppMater}.
	Magnon-pair formation is considered 
	via the resulting energy gap in $\chi_{\rm mag}^{-+}$ whose low-energy form
	at finite temperatures was determined within the framework of practical approximation.

	\begin{figure}[t]
	\begin{center}
	\includegraphics{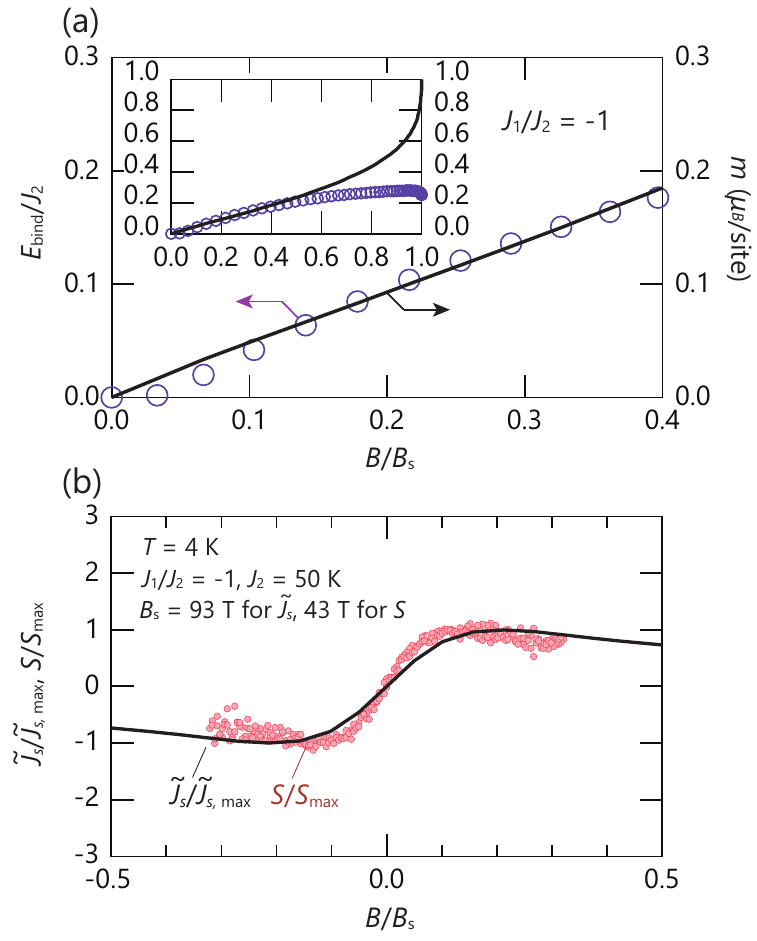}
	\end{center}
	\caption{
	(a) $B$ dependences of the magnon-pair binding energy $E_\text{bind}$ 
	and the calculated magnetic moment per site $m=2\langle S_j^z\rangle$ for a 1D frustrated 
	spin chain with $J_1/J_2 = -1$ \textcolor{black}{[see also Eq. (\ref{eq:J1-J2})]}~\cite{Sato13}. 
	%The g-factor is set to 2; thus the saturated $m$ is unity in units of $\mu_\text{B}$.
	%when converting the spin angular momentum 
	%into $m$. $B_\text{s}$ denotes the saturation field.
	%\tr{indicating that the saturated value of the magnetization per one site, $m_s$, is unity. 
	The inset shows the $B$ dependences up to $B/B_\text{s} = 1$
	\textcolor{black}{with $B_\text{s}$ being the saturation field}.
	(b) $B$ dependence of the calculated spin current $\tilde{J}_\text{s}$ 
	injected into a metal by single magnons which have an energy gap equal to 
	the magnon-pair binding energy. 
	%\textcolor{red}{\textit{reduced by 10} The detailed calculation of $\tilde{J}_\text{s}$
	%is given in Supplemental Material.} 
	The $B$ dependence of $S$ is also shown as data points for comparison.
	\textcolor{black}{$B$ is normalized by $B_\text{s} = 93$ T for $\tilde{J}_\text{s}$,
	calculated with $J_1/J_2 = -1$ and $J_2 = 50$ K while by $B_\text{s}$ $\textcolor{black}{\sim 43}$ T 
	for $S$~\cite{Svistov,Orlova}.}
	$\tilde{J}_\text{s}$ and $S$ are respectively normalized by their maximum values 
	$\tilde{J}_\text{s, max}$ and $S_\text{max}$.
	%\textcolor{black}{
	%We note that the theoretical $B_\text{s}$ is changed easily by changing $J_1$ and $J_2$~\cite{Hikihara}
	%in the spin-nematic TLL state while the $B$-linearity of $E_\text{bind}$ is not~\cite{Sato13}.
	%Thus, the $B$ dependence of $\tilde{J}_\text{s}$ changes little while changing $B_\text{s}$.}
	}
	\label{fig3}
	\end{figure}

	In Fig. 3(b), we show the \textcolor{black}{$B$ dependences} 
	of \textcolor{black}{calculated} $\tilde{J}_\text{s}$ \textcolor{black}{and measured $S$} at $T=4$ K 
	normalized by \textcolor{black}{their maximum values}.
	\textcolor{black}{We set $J_1/J_2 = -1$ and $J_2 = 50$ K in the calculation
	and normalized $B$ by the saturation field $B_\text{s}$ (see also the caption of Fig. 3).}
	\textcolor{black}{$\tilde{J}_\text{s}$ and $S$ increase} linearly with $B$ 
	near zero magnetic field. 
	This can be attributed to the growth of the uniform ferromagnetic moment 
	%and the resulting single magnons' angular momentum. 
	\textcolor{black}{and the angular momentum along $B$
	per single magnon~\cite{footnote_renormalizationFactor}.}
	Most importantly, $\tilde{J}_\text{s}$ 
	starts to be suppressed upon a further increase of $B$,
	and exhibits a broad peak structure around $|B| = 9$ T,
	capturing the marked feature \textcolor{black}{of $S$} observed experimentally.
	Since applying $B$ of this magnitude yields $E_\text{bind} \sim 3$ K
	[see also Fig. 3(a)], the $B$-induced suppression at $T = 4$ K can be ascribed to 
	a decrease in thermally excited single magnons that is induced by magnon-pair formation.
	\textcolor{black}{We stress that 
	the theoretical $B_\text{s}$ is varied easily by changing $J_1$ and $J_2$~\cite{Hikihara}
	in the spin-nematic TLL state while the $B$-linearity of $E_\text{bind}$ is not~\cite{Sato13}.
	Thus, the $B$ dependence of $\tilde{J}_\text{s}$ little depends on change of $B_\text{s}$.
	This indicates that a difference between the theoretical $B_\text{s} = 93$ T 
	and the experimental $B_\text{s} \sim 43$ T is not essential in reproducing 
	the characteristic $B$-dependence of $S$.}
	
	We note that for LiCuVO$_4$, the 3D spin spiral correlation likely coexists 
	with the magnon-pair one above magnetic ordering temperatures $\sim 3$ K [see also Fig. 1(c)].
	%we indeed found that the slope of $S$ vs $B$ around the origin saturates towards $\sim 3$ K 
	%in magnitude (see Supplemental Material). 
	Since $B$ is applied parallel to the spiral axis along the $c$-axis, the low-$B$ SSE 
	is similar to antiferromagnetic SSEs in canted phases~\cite{Seki, Wu2016}.
	In these previous cases, spin-Seebeck coefficients exhibit positive sign along with 
	the same $B$ dependences as those of $M$. These features are expected to be embedded 
	in our low-$B$ SSE results. Integrating such effects into the above calculation 
	will yield a more quantitative result 
	while the $B$-induced suppression of magnon-pair origin should carry over.%~\cite{footnote_spiralXZeeman2}.	

	\begin{figure}[b]
	\begin{center}
	\includegraphics{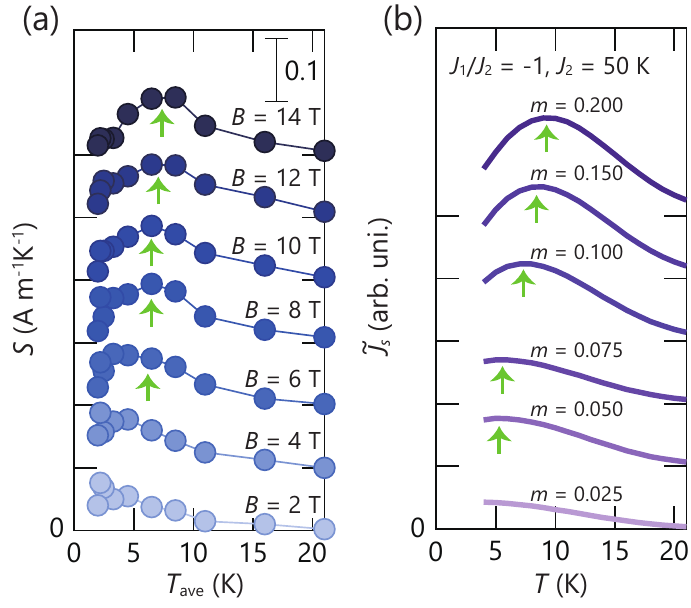}
	\end{center}
	\caption{
	(a) $T_\text{ave}$ dependence of $S$ at several $B$. 
	\textcolor{black}{Datasets are shifted by multiples of 0.1 $\text{A}\text{m}^{-1}	\text{K}^{-1}$.}
	(b) $T$ dependence of the calculated spin current $\tilde{J}_\text{s}$ that is
	%injected from a spin-nematic Tomonaga-Luttinger liquid to a metal.
	injected into a metal by single magnons with an energy gap equal to the 
	magnon-pair binding energy. %of a spin-nematic Tomonaga-Luttinger liquid.
	$m=2\langle S_j^z\rangle$ is the \textcolor{black}{magnetic moment} 
	per site.
	%\textcolor{red}{\textit{reduced by 10}
	%The detailed calculation of $\tilde{J}_\text{s}$ is described in Supplemental Material.}
	%$m$ denotes calculated magnetic moment in units of $\mu_\text{B}$ per site. 
	%See also the caption of Fig. 3 for the definition of $m$.
	}
	\end{figure}

	In Fig. 4, we compare the $T_\text{ave}$ dependences of $S$ for several $B$ with 
	our theoretical calculations, in which finite-temperature effects on the single-magnon dynamics
	are considered besides the magnon-pair binding energy.
%When $B = 2$ T is applied, $S$ increases while decreasing $T_\text{ave}$ from 20 K, 
%and only saturates towards low $T_\text{ave}$.
%However, as $B$ is increased, a broad peak structure shows up, reflecting 
%the $B$-induced suppression of $S$. 
	When $B$ is below $\sim 5$ T, $S$ only saturates toward low $T_\text{ave}$ 
	as seen in Fig. 4(a). However, when $B$ is above $\sim 5$ T, a broad peak structure emerges,
	and its peak position gradually shifts from $\sim 5$ K to $\sim 8$ K while increasing $B$ to 14 T.
	These temperature dependences are also successfully captured by our calculation 
	based on formula~(\ref{eq_SpinCurrent2}), as shown in Fig. 4(b).
	\textcolor{black}{This shows that the broad peaks stem from the competition between 
	a decrease in the single-magnon density due to the magnon-pair formation and
	an increase in the single-magnon lifetime at low temperatures.
	Additionally, the agreement between Figs. 4(a) and (b) indicates that 
	the peak shift caused by increasing $B$ could be attributed to 
	an increase in the angular momentum along $B$ per single magnon~\cite{footnote_renormalizationFactor}:
	%subject to the magnon-pair formation:
	Such increased angular momentum enhances SSE at high temperature where 
	%as long as 
	the $B$-induced magnon-pair binding energy can be overcome by thermal fluctuation;
	otherwise, SSE is decreased more greatly toward low temperature via magnon-pair formation.
	This can be responsible for the peak shift observed in Fig. 4(a).
	}
	\textcolor{black}{Overall, the agreement between the experimental and theoretical results shows that 
	the $B$ and $T$ dependences of $S$ can be well explained by magnon-pair formation.
	We also note that our results point to exchange of spin-1 as the most relevant magnetic 
	interaction at the interface in SSE.}

     %\tr{These agreements between the experiment and the theory clearly show that the unusual behavior of the SSE 
     %is indeed attributed to the realization of a spin-nematic TLL and the magnon-pair formation. They also indicate 
     %that the most relevant magnetic interaction at the interface is the spin-1 exchange.} 

%%%%%%%%%%%%%%%%%%%%%%%%%%%%%%%%%%%%%%%%%%%%
%%%%%%%%%%%%%%%%%%%%%%%%%%%%%%%%%%%%%%%%%%%%
%%%%%%%%%%%%%%%%%%%%%%%%%%%%%%%%%%%%%%%%%%%%
%\noindent
\textit{Summary}$-$
	We \textcolor{black}{observed} the magnetic-field-induced suppression 
	of the SSE in \textcolor{black}{a} quasi-1D frustrated spin-chain system LiCuVO$_4$,
	an established model material \textcolor{black}{for} the spin-nematic correlation. 
	%The unusual suppression of the SSE was observed to appear in spite of the field-linear magnetization curves. 
	A broad peak structure was also found to appear in the temperature dependence of the spin-Seebeck voltage, 
	and to shift toward high temperatures while increasing magnetic field. 
    These experimental results were well reproduced by a microscopic calculation of 
	the interfacial spin current where the magnon-pair binding energy
	\textcolor{black}{and its resulting} energy gap of the single magnons 
    are taken into consideration.
	%Additionally, a broad peak structure was found to appear 
	%in the temperature dependence of the spin-Seebeck voltage, 
	%and to shift towards high temperatures while increasing magnetic field. 
	%This feature was also captured by the computed spin current  
	%that \textcolor{black}{accounts} for the competition between the energy gap 
	%\textcolor{black}{due to magnon-pair binding and the prolonged lifetime of single magnons}. 
	Our result indicates that SSE is a powerful tool for detecting signatures of spin-nematic states 
	and their transport properties.   

	%To our knowledge, this study is the first to detect a signature of the spin-nematic
	%]correlation in transport experiment. 
%%%%%%%%%%%%%%%%%%%%%%%%%%%%%%%%%%%%%%%%%%%%
%%%%%%%%%%%%%%%%%%%%%%%%%%%%%%%%%%%%%%%%%%%%
%%%%%%%%%%%%%%%%%%%%%%%%%%%%%%%%%%%%%%%%%%%%
%\noindent\textit{Note added.} 

%%%%%%%%%%%%%%%%%%%%%%%%%%%%%%%%%%%%%%%%%%%
%%%%%%%%%%%%%%%%%%%%%%%%%%%%%%%%%%%%%%%%%%%
%%%%%%%%%%%%%%%%%%%%%%%%%%%%%%%%%%%%%%%%%%%
%\noindent
\textit{Acknowledgments}
We thank \textcolor{black}{Toshiya} Hikihara for fruitful discussion on the magnon-pair binding energy.
\textcolor{black}{We also thank Takashi Kikkawa for experimental assistance.}
This work is supported by JSPS (KAKENHI No. 17H04806, No. 18H04215, No. 18H04311, 
\textcolor{black}{No. 18H05854} and No. 26247058 and the Core-to-Core program 
``International research center for new-concept spintronics devices") and MEXT 
(Innovative Area ``Nano Spin Conversion Science" (No. 26103005)). 
D. H. was supported by the Yoshida Scholarship Foundation through the Doctor 21 program. 
M. S. was supported by Grant-in-Aid for Scientific Research on Innovative Area, 
``Nano Spin Conversion Science" (Grant No.17H05174)
\textcolor{black}{
and ``Quantum Liquid Crystals" (Grant No.19H05825)
as well as} JSPS KAKENHI (Grant No. 17K05513 and No. 15H02117).

%%%%%%%%%%%%%%%%%%%%%%%%%%%%%%%%%%%%%%%%%%%
%%%%%%%%%%%%%%%%%%%%%%%%%%%%%%%%%%%%%%%%%%%
%%%%%%%%%%%%%%%%%%%%%%%%%%%%%%%%%%%%%%%%%%%

\end{document}